\documentclass[12pt,twoside]{article}

\usepackage{fleqn,epsfig,times}

\newcommand{\stxt}[1]{\mbox{\scriptsize #1}}

\newcommand{\MW}{M_{\stxt{W}}}
\newcommand{\MZ}{M_{\stxt{Z}}}
\newcommand{\MH}{M_{\stxt{H}}}
\newcommand{\OMT}[1]{\overline{M}_{\hspace{-.3ex}\stxt{#1}}}
\newcommand{\OMP}{\overline{M}}
\newcommand{\OGP}{\overline{\Gamma}}
\newcommand{\me}{m_{\stxt{e}}}
\newcommand{\mt}{m_{\stxt{t}}}
\newcommand{\sw}{s_{\mbox{\tiny{W}}}}
\newcommand{\cw}{c_{\mbox{\tiny{W}}}}
\newcommand{\dsw}{\delta s_{\mbox{\tiny{W(2)}}}}
\newcommand{\GF}{G_{\stxt{F}}}
\newcommand{\as}{\alpha_{\stxt{s}}}

\newcommand{\rt}[1]{\left(#1\right)^{\frac{1}{2}}}
\newcommand{\irt}[1]{\left(#1\right)^{-\frac{1}{2}}}

\newcommand{\lesim}{\raisebox{-.3ex}{$_{\textstyle <}\atop^{\textstyle\sim}$}}

\newcommand{\xia}{\xi_1^{\gamma}}
\newcommand{\xiaz}{\xi^{\gamma\stxt{Z}}}
\newcommand{\xiza}{\xi^{\stxt{Z}\gamma}}
\newcommand{\xiz}{\xi_1^{\stxt{Z}}}
\newcommand{\xizz}{\xi_2^{\stxt{Z}}}
\newcommand{\xiw}{\xi_1^{\stxt{W}}}
\newcommand{\xiww}{\xi_2^{\stxt{W}}}

\newcommand{\eqref}[1]{(\ref{#1})}

\newdimen\mybls                    
\mybls=\baselineskip               
\advance\mybls -1ex                
\newcounter{address}
\def\theaddress{\alph{address}}
\def\mymakeadmark#1{\hbox{$^{\rm #1}$}}

\def\address#1{\addressmark\begingroup
  \xdef\mytempa{\theaddress}\let\\=\relax
  \def\protect{\noexpand\protect\noexpand}\xdef\myaddress{\myaddress
  \protect\addresstext{\mytempa}{#1}}\endgroup}
\def\myaddress{}

\def\addressmark{\stepcounter{address}%
  \xdef\mytempa{\theaddress}\mymakeadmark{\mytempa}}

\def\addresstext#1#2{\leavevmode \begingroup
  \small \it \mymakeadmark{#1}#2\par \endgroup
  \vskip\mybls}

\title{ \begin{flushright}\normalsize
	CERN--TH/2000--207\\
	DESY 00--101\\
	KA--TP--15--2000\\[6ex]
	\end{flushright}
	Calculation of fermionic two-loop contributions \\ to muon
        decay\thanks{Talk given by A. F. at the 5th Zeuthen Workshop on Elementary
        Particle Theory "Loops and Legs in Quantum Field Theory",
        Bastei/K\"onigstein, Germany, April 9--14, 2000}}

\author{A. Freitas\address{Deutsches Elektronen-Synchrotron DESY, Notkestr.
        85, \\ D-22603 Hamburg, Germany}\xdef\desy{\theaddress},
        S. Heinemeyer\hbox{$^{\rm \desy}$},
        W. Hollik\address{Institut f\"ur Theoretische Physik, Universit\"at
        Karlsruhe, \\ D-76128 Karlsruhe, Germany}\xdef\ka{\theaddress},
        W. Walter\hbox{$^{\rm \ka}$},
        G. Weiglein\address{Theory Division, CERN, CH-1211 Geneva 23,
        Switzerland}}

\date{\vskip\mybls \myaddress}

\begin{document}

\maketitle

\begin{abstract} The computation of the correction $\Delta r$ in the
W-Z mass correlation, derived from muon decay, is described at the two-loop level
in the Standard Model. Technical aspects which become relevant at this level
are studied, e.g. gauge-parameter independent mass renormalization, 
ghost-sector renormalization and the treatment
of $\gamma_5$. Exact results for $\Delta r$ and the W mass prediction
including ${\mathcal{O}}(\alpha^2)$ corrections with fermion loops are
presented and compared with previous results of a next-to-leading order
expansion in the top-quark mass. \end{abstract}

\thispagestyle{empty}

\setcounter{page}{0}
\setcounter{footnote}{0}

\newpage

\section{Introduction}

The electroweak Standard Model (SM), together with the theory of strong
interactions (Quantum Chromodynamics, QCD), provides a 
comprehensive description of 
experimental data with remarkable consistency. In order to further
test the validity of the SM predictions and restrict its only missing
parameter, the Higgs boson mass, precision measurements play a key role. The
analysis of precision observables is sensitive to quantum corrections in the
theoretical predictions, which depend on all the parameters of the model. 
In this way, 
the top-quark mass, $\mt$, had been predicted in the region where it was
experimentally found.

The constraints on the Higgs mass, $\MH$, are still rather weak 
since $\MH$ appears
only logarithmically in the leading order SM predictions. 
Therefore it is of high
interest to further reduce experimental and theoretical uncertainties. An
important quantity in this context is the quantum correction $\Delta r$  in the
relation of the gauge boson masses $\MW, \MZ$ with the Fermi constant $\GF$ and
the fine structure constant $\alpha$ \cite{S:dr}.

This relation is established in terms of the muon decay width, which was first
described in the Fermi Model as a four-fermion interaction $\mu^- \rightarrow
e^-\,\nu_\mu \, \bar{\nu}_e$ with coupling
$\GF$. This yields for the muon decay width
\begin{equation}
\Gamma_\mu = \frac{\GF^2 \, m_\mu^5}{192 \pi^3} \;
F\left(\frac{\me^2}{m_\mu^2}\right) \left(1 + \Delta q \right), \label{GFdef}
\end{equation}
where $F(\me^2/m_\mu^2)$ subsumes effects of the electron mass on the 
final-state phase space and $\Delta q$ denotes the QED corrections in the 
Fermi Model.
Including two-loop QED corrections
\cite{RS:QED2,SS:QED2} one obtains from the measurement of the muon decay
width \cite{C:pdg} $\GF = (1.16637 \pm 0.00001)\, 10^{-5} \mbox{ GeV}^{-2}$.

The decay process in the SM, on the other side,
involves the exchange of a W boson
and additional electroweak higher order corrections. Although not part of the
Fermi Model, tree-level W propagator effects are conventionally included in eq.
\eqref{GFdef} by means of a factor $(1+\frac{3}{5} m_\mu^2/\MW^2)$. Yet, this
is of no numerical significance. The relation between the Fermi constant $\GF$
and the SM parameters is expressed as
\begin{equation}
\frac{\GF}{\sqrt{2}} =
 \frac{e^2}{8 \sw^2 \MW^2} \, \left(1+ \Delta r \right). \label{drdef}
\end{equation}
Here the SM radiative corrections are included in $\Delta r$ (with $\sw^2 =
1-\MW^2/\MZ^2$).

Since $\GF$ is known with high accuracy it can be taken 
as an input parameter in
equation \eqref{drdef} in order to obtain a prediction for the W mass
($\MW$ is still afflicted with a considerable experimental
error, $\MW = 80.419 \pm  0.038$ GeV \cite{LEPEWWG:Wmass}):
\begin{equation}
\MW^2 = \MZ^2 \left[ \frac{1}{2} + \sqrt{\frac{1}{4}
 -\frac{\alpha\pi}{\sqrt{2}\GF \MZ^2}
  \left(1+\Delta r \right)} \, \right] \hspace{-.5ex}. \label{MWpred}
\end{equation}
Since $\Delta r$ itself depends on $\MW$, eq. \eqref{MWpred} is to be
understood as an implicit expression.

One expects substantial improvement on the W mass determination from future
colliders, which will allow for increasingly stringent constraints on the
SM from the comparison of the prediction with the experimental
value for $\MW$. This requires an accurate theoretical determination of 
the quantity $\Delta r$.

The one-loop result \cite{S:dr,SM:dr1loop} can be split into the following
contributions:
\begin{equation}
\Delta r^{(\alpha)} = \Delta\alpha - \frac{\cw^2}{\sw^2} \Delta\rho + \Delta
r_{\stxt{rem}}(\MH).
\end{equation}
Dominant corrections arise from the shift in the fine-structure constant,
$\Delta\alpha$, due to large logarithms of light-fermion masses 
($\approx 6\%$), and
from the leading contribution $\propto \mt^2$ to the $\rho$ parameter 
resulting from the top/bottom doublet, which 
enters through $\Delta\rho$ ($\approx 3.3\%$). The full $\MH$--dependence is
contained in the remainder $\Delta r_{\stxt{rem}}$ ($\approx 1\%$).
  
Furthermore, QCD corrections of ${\mathcal{O}}(\alpha\as$) \cite{QCD1} and
${\mathcal{O}}(\alpha\as^2$) \cite{QCD2} are known. Leading electroweak
fermionic ${\mathcal{O}}(\alpha^2)$ contributions were first taken into account
by means of resummation relations \cite{resum}. Different approaches for the
calculation of electroweak two-loop corrections have been pursued, involving
expansions for large values of $\MH$ \cite{BV:mhexp} and $\mt$
\cite{BH:mtlead,mtquad,mtsqr}.

It turned out that both the leading ${\mathcal{O}}(\alpha^2\mt^4/\MW^4)$
\cite{mtquad} and the next-to-leading ${\mathcal{O}}(\alpha^2\mt^2/\MW^2)$
\cite{mtsqr} coefficients in the $\mt$ expansion yield important corrections of
comparable size. Therefore a complete two-loop calculation of fermionic
contributions would be desirable in order to further reduce the theoretical
uncertainty, in particular if one considers that these contributions are
dominant already in the one-loop result.

A first step into this direction was the determination of the exact Higgs mass
dependence of the fermionic ${\mathcal{O}}(\alpha^2)$ corrections to $\Delta r$
\cite{BW:MHdep}. Recently the full calculation of these contributions
\cite{mudec2} has been accomplished. The results include all electroweak
two-loop diagrams with one or two fermion loops without any expansion in $\mt$
or $\MH$. This talk gives an overview on the techniques and the results of this
calculation.

\section{Calculational methods}

Since in the present calculation all possibly infrared divergent photonic 
corrections are already
contained in the definition \eqref{GFdef} of the Fermi constant $\GF$ and mass
singularities are absorbed in the running of the electromagnetic coupling, 
$\MW$ represents the scale for the electroweak corrections in $\Delta
r$. Therefore it is possible to neglect all fermion masses except the top
quark mass and the momenta of the external leptons so that the 
muon decay diagrams reduce to vacuum diagrams.

All QED contributions to the Fermi Model have to be excluded in the computation
of $\Delta r$ as they are separated off in the definition of $\GF$, see eq.
\eqref{GFdef}. When writing the factor $(1+\Delta q)$ formally as
$(1+\Delta\omega)^2$ the decay width 
up to ${\mathcal{O}}(\alpha^2)$ can be decomposed as
\begin{equation}
\setlength{\arraycolsep}{2pt}
\begin{array}{rcl}
\Gamma_\mu &=& \Gamma_{\mu,\stxt{tree}} (1+\Delta\omega)^2 \, (1+\Delta r)^2
\\[1ex]
&=& \Gamma_{\mu,\stxt{tree}}
 \Bigl[
  1 + 2(\Delta\omega^{(\alpha)} + \Delta r^{(\alpha)})
+ (\Delta\omega^{(\alpha)} + \Delta r^{(\alpha)})^2
+ 2 \Delta\omega^{(\alpha)} \Delta r^{(\alpha)} \\[1ex]
&& \qquad\quad + 2 \Delta\omega^{(\alpha^2)}
+ 2 \Delta r^{(\alpha^2)} + {\mathcal{O}}(\alpha^3) \Bigr].
\end{array}
\end{equation}
Apart from the one-loop contributions this includes two-loop QED corrections
$\Delta\omega^{(\alpha^2)}$ and mixed contributions of QED and weak corrections
$\Delta\omega^{(\alpha)} \Delta r^{(\alpha)}$ which both thus have to be
excluded in $\Delta r^{(\alpha^2)}$. For fermionic two-loop diagrams it is 
possible to find a one-to-one correspondence between QED graphs in Fermi Model
and SM contributions.

The renormalization is performed in the on-shell scheme. In this
context the mass renormalization constants require the computation of two-loop
self-energy diagrams with non-vanishing momentum.

All decay amplitudes and counterterm contributions have
been generated with the program \emph{FeynArts 2.2} \cite{feynarts}. The
amplitudes are algebraically reduced by means of a general tensor integral
decomposition for two-loop two-point functions with the program \emph{TwoCalc}
\cite{twocalc}, leading to a fixed set of standard scalar integrals. Analytical
formulae are known for the scalar one-loop \cite{HV:oneloop} and two-loop
\cite{DT:twovac} vacuum integrals whereas the two-loop self-energy diagrams can
be evaluated numerically by means of one-dimensional integral
representations \cite{twoself}.

In order to apply an additional check the calculations were performed within a
covariant $R_\xi$ gauge, which introduces one gauge parameter $\xi_i, i =
\gamma, \mbox{Z,W}$, for each gauge boson. It has been explicitly checked at the
algebraic level that the gauge parameter dependence of the final result drops
out.

\section{On-shell renormalization}

For the determination of the one-loop counterterms (CTs) and renormalization
constants the conventions of  ref. \cite{D:habil} are adopted. 
Two-loop renormalization constants 
enter via the counterterms for the transverse W propagator and the charged
current vertex:
\begin{eqnarray}
\left[
\mbox{\raisebox{-1mm}{\psfig{figure=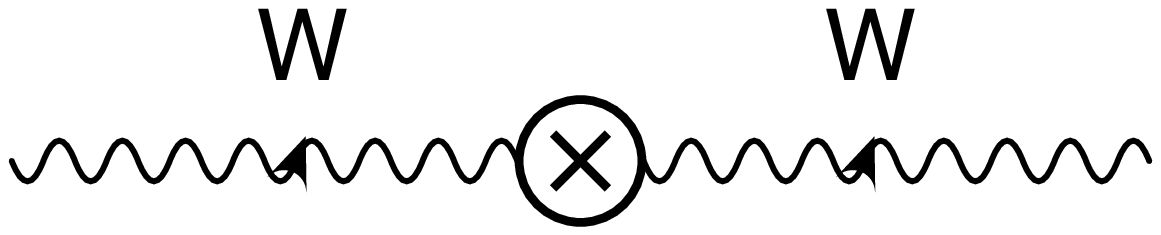,width=25mm}}}
\right]_{\stxt{T}}
 &=&
 \delta Z_{(2)}^{\stxt{W}}(k^2-\MW^2) - \delta M^2_{\stxt{W(2)}}-
        \delta Z_{(1)}^{\stxt{W}} \delta M^2_{\stxt{W(1)}}, \label{w-ct} \\[2ex]
\mbox{\raisebox{-8mm}{\psfig{figure=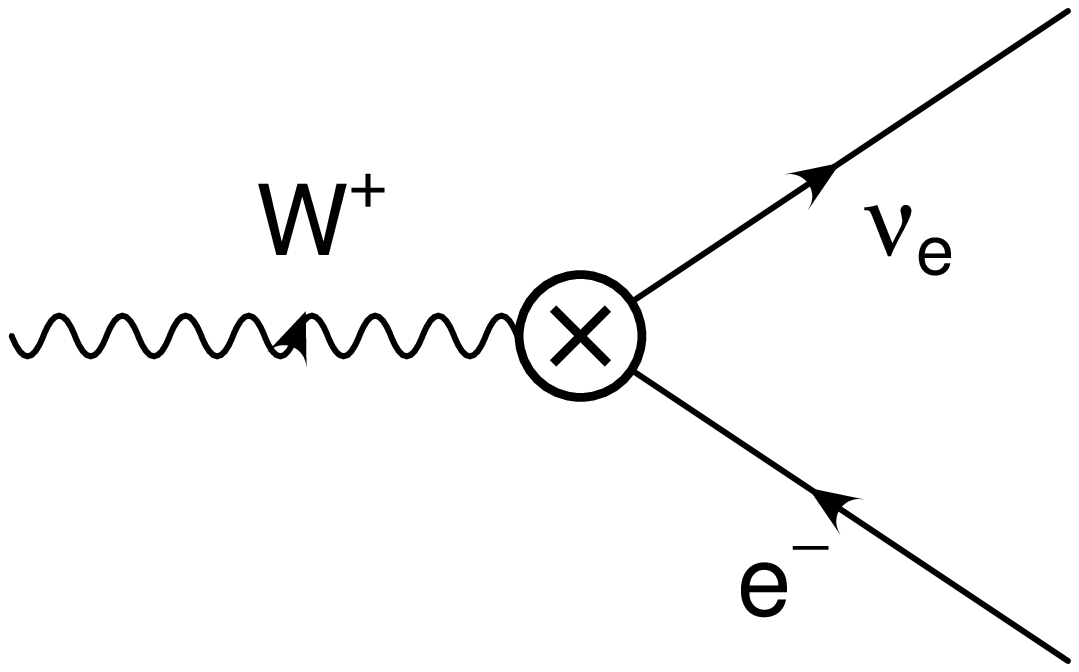,width=25mm}}}
 &=& i \frac{e}{\sqrt{2} \sw} \, \gamma_\mu \omega_-
  \biggl[ \delta Z_{e(2)}
  - \frac{\dsw}{\sw}
  +\frac{1}{2} \left(\delta Z_{(2)}^{e\stxt{L}}+\delta Z_{(2)}^{\stxt{W}} +
        \delta Z_{(2)}^{\nu \stxt{L}} \right) \\[-1ex]
 &&+ \; \mbox{(1-loop renormalization constants)} \biggr]. \nonumber
\end{eqnarray}
Here, $\delta Z^\phi$ denotes the field-renormalization constant 
of the field $\phi$, $\delta
M^2_\phi$ the corresponding mass CT, and $\delta Z_e$ the 
charge-renormalization constant. The numbers
in parentheses indicate the loop order. The mixing angle CT, $\dsw$, is
expressible through the gauge boson mass CTs. Throughout
this paper, the two-loop contributions always include the subloop
renormalization.

The on-shell masses are defined as the position of the propagator poles. 
Starting at the two-loop level, 
it has to be taken into account that there is a difference
between the definition of the mass $\widetilde{M}^2$ as the real pole of the
propagator $D$,
\begin{equation}
\Re \left\{ D_{\stxt{T}})^{-1}({\widetilde{M}}^2) \right\}
 = 0,
\end{equation}
and the real part $\overline{M}^2$ of the complex pole,
\begin{equation}
(D_{\stxt{T}})^{-1}({\mathcal{M}}^2) = 0, \qquad
{\mathcal{M}}^2 = \OMP^2  - i \OMP\, \OGP.
\label{complex}
\end{equation}
The imaginary part of the complex pole is associated with the width
$\overline{\Gamma}$. The definition (\ref{complex})
yields for the W mass CT
\begin{equation}
\delta \OMT{W(2)}^2 = \Re \bigl\{
  \Sigma_{\stxt{T(2)}}^{\stxt{W}}(\OMT{W}^2) \bigr\}
 -  \delta Z_{(1)}^{\stxt{W}} \delta \OMT{W(1)}^2
 + \Im\bigl\{\Sigma_{\stxt{T}(1)}^{\stxt{W}^/}(\OMT{W}^2)\bigr\}
     \:
   \Im\bigl\{\Sigma_{\stxt{T}(1)}^{\stxt{W}}(\OMT{W}^2)\bigr\}, \label{MWct}
     \nonumber
\end{equation}
whereas for the real pole definition the last term of eq.
\eqref{MWct} is missing. $\Sigma_{\stxt{T}}^{\stxt{W}},
\Sigma_{\stxt{T}}^{\stxt{W}^/}$  denote the transverse W self-energy and its
momentum derivative. Similar expressions hold for the Z boson.

The W and Z mass CTs determine 
the two-loop mixing angle CT, $\dsw$,
which has to be gauge invariant since $\sw$ is an observable
quantity. With the use of a general $R_\xi$ gauge it could be
explicitly checked that $\dsw$
is gauge-parameter independent for the complex-pole mass definition,
whereas the real-pole definition leads to a gauge dependent $\dsw$.
This is in accordance with the expectation from S-matrix theory
\cite{compole}, where
the complex pole represents a gauge-invariant mass definition.

It should be noted that the mass definition via the complex pole corresponds to
a Breit-Wigner parameterization of the resonance shape with a constant width.
For the experimental determination of the gauge boson masses, however,
a Breit-Wigner ansatz with a running width is used. This has to be 
accounted for
by a shift of the values for the complex pole masses 
\cite{BLRS:pole},
\begin{equation}
\OMP = M - \frac{\Gamma^2}{2 M},
\end{equation}
which yields the relations
\begin{eqnarray}
\begin{array}{lcl}
 \OMT{Z} & = & \MZ - 34.1 \mbox{ MeV,}  \\
 \OMT{W} & = & \MW - 27.4\; (27.0) \mbox{ MeV} \quad 
  \mbox{ for } \quad \MW = 80.4\; (80.2) \mbox{ GeV.}
\end{array}
\end{eqnarray}
For $\MZ$ and $\Gamma_{\stxt{Z}}$ the experimental numbers are taken.
The W mass is a calculated quantity, 
and therefore also a theoretical value for the W boson
width should be applied here.\footnote{In the version of this paper appearing in
the proceedings for simplicity a fixed shift of leading order has been used.}
The results above are obtained from the approximate, but sufficiently
accurate expression for the W width, 
\begin{equation}
\Gamma_{\stxt{W}} = 3 \frac{\GF \MW^3}{2 \sqrt{2} \pi} \,
 \left( 1 + \frac{2 \as}{3 \pi} \right) .
\end{equation}

\vspace{1ex}
At the subloop level, also the Faddeev-Popov ghost sector
has to be renormalized.
The gauge-fixing sector for the gauge fields
$A^\mu$, $Z^\mu$, $W^{\pm\mu}$ ($\chi$, $\phi^\pm$  denote the
unphysical Higgs scalars)
\[
{\mathcal{L}}_{\stxt{gf}} = -\frac{1}{2}
  \Big((F^\gamma)^2+(F^{\stxt{Z}})^2+F^+F^-+F^-F^+\Big),
\]
\vspace{-2.5ex}
\begin{equation}
F^\gamma =\irt{\xia} \partial_\mu A^\mu +\frac{\xiaz}{2} \partial_\mu Z^\mu,
\label{gaugefix}
\end{equation}
\vspace{-2ex}
\[
F^{\stxt{Z}} = \irt{\xiz} \partial_\mu Z^\mu 
        + \frac{\xiza}{2} \partial_\mu A^\mu
        - \rt{\xizz} \MZ \, \chi,
\]
\vspace{-2.5ex}
\[
F^\pm =\irt{\xiw}\partial_\mu W^{\pm\mu} \,\mp\,i\rt{\xiww} \MW
 \, \phi^\pm
\]
does not need renormalization.
Accordingly, one can either introduce the gauge-fixing term 
after renormalization or 
renormalize the gauge parameters in
such a way that they compensate the renormalization of the fields and masses. 
Both methods ensure that no counterterms arise from the gauge-fixing sector but
they differ in the treatment of the ghost Lagrangian, 
which is given by the variation of the
functionals $F^a$ under infinitesimal gauge transformations
$\delta\theta_b$,
\begin{equation}
{\mathcal{L}}_{\stxt{FP}} = \sum_{a,b = \gamma, \stxt{Z}, \pm} \bar{u}^a \,
  \frac{\delta F^a}{\delta \theta^b} \, u^b. \label{FP}
\end{equation}
In the latter case, which was applied in this work, 
additional counterterm contributions for the ghost sector arise from the
gauge parameter renormalization.
The parameters $\xi^a_i$ in \eqref{gaugefix} are renormalized 
such that their CTs $\delta\xi^a_i$ exactly cancel the
contributions from the renormalization of the fields and masses and that the
renormalized gauge parameters comply with the $R_\xi$ gauge.


\section{Treatment of the $\gamma_5$--problem}

In four dimensions the algebra of the $\gamma_5$--matrix is defined by the two
relations
\begin{eqnarray}
\bigl\{ \gamma_5, \gamma_\alpha \bigr\} = 0 \qquad \mbox{for} \qquad \alpha =
  1, \dots, 4 \label{anticomm} \\[1ex]
\mbox{Tr} \bigl\{ \gamma_5 \gamma^\mu\gamma^\nu\gamma^\rho\gamma^\sigma \bigr\}
  = 4i \epsilon^{\mu\nu\rho\sigma}\mbox{.} \label{trg4}
\end{eqnarray}
It is impossible to translate both relations simultaneously into $D
\neq 4$ dimensions without encountering inconsistencies \cite{HV:dreg}.

A certain treatment of $\gamma_5$ might break symmetries, i.e. violate
Slavnov-Taylor (ST) identities which would have to be restored with extra
counterterms. Even after this procedure a residual scheme dependence can
persist which is associated with $\epsilon$-tensor expressions originating from
the treatment of \eqref{trg4}. Such expressions cannot be canceled by
counterterms. If they broke ST identities this would give rise to anomalies.

't Hooft and Veltman \cite{HV:dreg} suggested a consistent scheme which was
formulated by Breitenlohner and Maison \cite{BM:G5scheme} as a separation of
the first four and the remaining dimensions of the $\gamma$-Matrices
(HVBM-scheme). It has been shown \cite{SMren} that the SM with HVBM
regularization is anomaly-free and renormalizable. This shows that
$\epsilon$-tensor terms do not get merged with divergences.

The naively anti-commuting scheme, which is widely used for one-loop
calculations, extends the rule \eqref{anticomm} to $D$ dimensions but abandons
\eqref{trg4},
\begin{eqnarray}
\bigl\{ \gamma_5, \gamma_\alpha \bigr\} = 0 \qquad \mbox{for} \qquad \alpha =
  1, \dots, D \\[1ex]
\mbox{Tr} \bigl\{ \gamma_5 \gamma^\mu\gamma^\nu\gamma^\rho\gamma^\sigma \bigr\}
  = 0.
\end{eqnarray}
This scheme is unambiguous but does not reproduce the four-dimensional case.

\begin{figure}[tb]
\centerline{
\psfig{figure=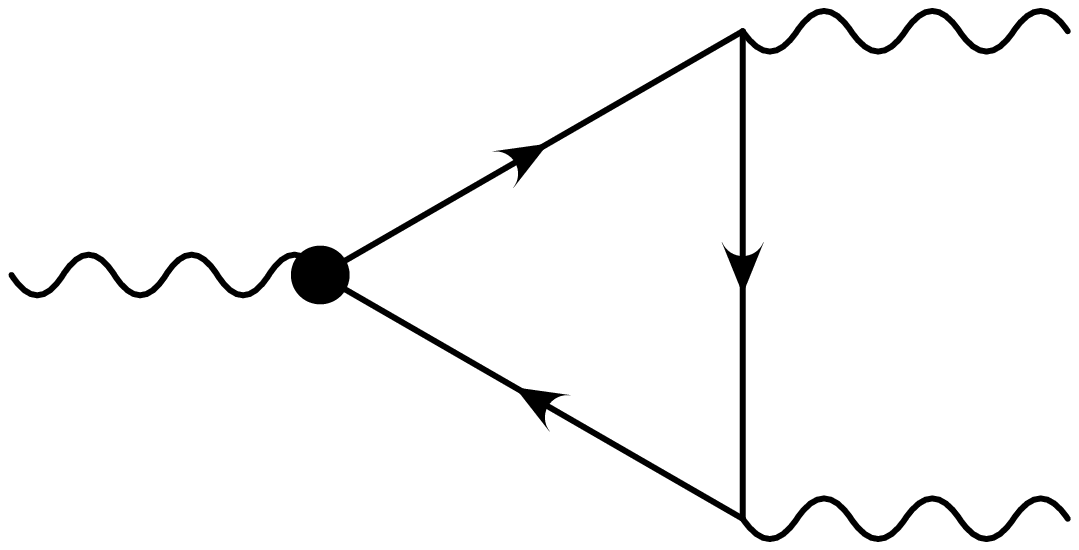,width=4cm}
\hspace{10mm}
\psfig{figure=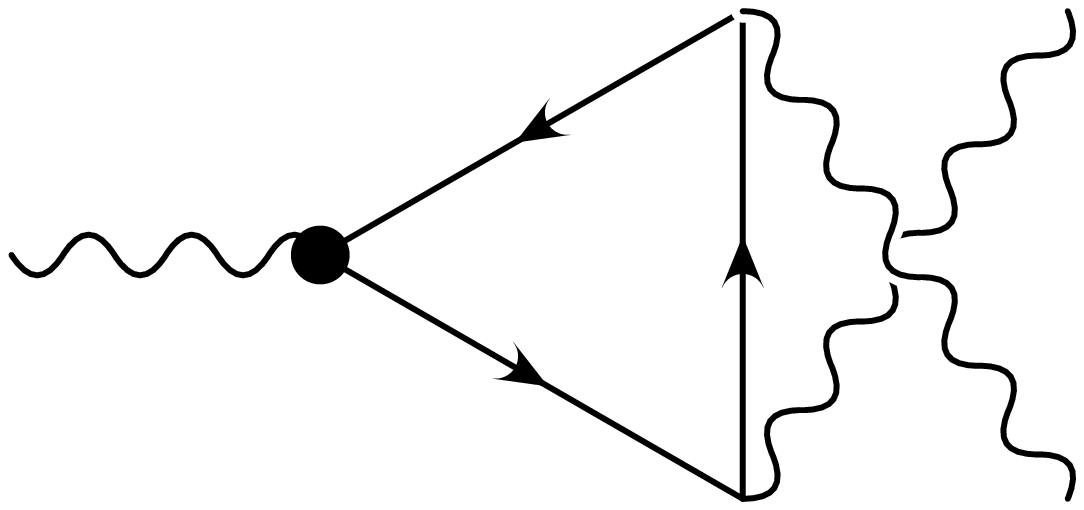,width=4cm}
}
\caption{Generic Standard Model triangle diagrams subject to the
$\gamma_5$--problem.}
\label{fig:tri}
\end{figure}
In the SM particularly triangle diagrams (Fig. \ref{fig:tri})
containing chiral couplings are sensitive to the $\gamma_5$--problem. For the
presented work one-loop triangle diagrams have been explicitly calculated in both
schemes. While the naive scheme immediately respects all ST
identities the HVBM scheme requires the introduction of additional finite
counterterms. Even after this procedure finite differences remain between the
results of the two schemes, showing that the naive scheme is
inapplicable in this case.

\begin{figure}[tb]
\centerline{
\psfig{figure=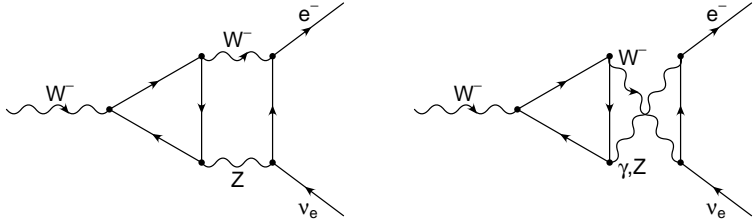,width=10cm}
}
\vspace{-2ex}
\caption{CC vertex diagrams with triangle subgraphs.}
\label{fig:CCvert}
\end{figure}
In the calculation of $\Delta r$ triangle diagrams appear as subloops of
two-loop charged current (CC) vertex diagrams (Fig. \ref{fig:CCvert}). One
finds that for the difference terms between both schemes this loop can be
evaluated in four dimensions without further difficulties. This can be
explained by the fact that renormalizability forbids divergent contributions to
$\epsilon$-tensor terms from higher loops in the HVBM scheme. The
$\epsilon$-tensor contributions from the triangle subgraph in the HVBM scheme
meet a second $\epsilon$-tensor term from the outer fermion lines in Fig.
\ref{fig:CCvert}, thereby resulting in a non-zero contribution to $\Delta r$.

Computations in the HVBM scheme can get very tedious because of the necessity
of additional counterterms. Therefore another method shall be examined. One
can consider a "mixed" scheme that uses both relations \eqref{anticomm} and
\eqref{trg4} in $D$ dimensions despite their mathematical inconsistency. This
scheme is plagued by ambiguities of ${\mathcal{O}}(D-4)$. 
When applied to the calculation of one-loop triangle diagrams the results
immediately respect all ST identities and differ from the HVBM
results only by terms of ${\mathcal{O}}(D-4)$,
\begin{equation}
\Gamma^{\stxt{HVBM}}_{\Delta (1)} =
\Gamma^{\stxt{mix}}_{\Delta (1)} + {\mathcal{O}}(D-4).
\end{equation}
Since for the difference
terms the second loop can be evaluated in four dimensions, this also holds for
the two-loop CC diagrams,
\begin{equation}
\Gamma^{\stxt{HVBM}}_{\stxt{CC(2)}} =
\Gamma^{\stxt{mix}}_{\stxt{CC(2)}} + {\mathcal{O}}(D-4).
\end{equation}

Thus the mixed scheme, despite being mathematically inconsistent,
can serve as a technically easy prescription for the correct calculation of the
CC two-loop contributions.

\section{Results}

\begin{figure}[tb]
\centerline{
\psfig{figure=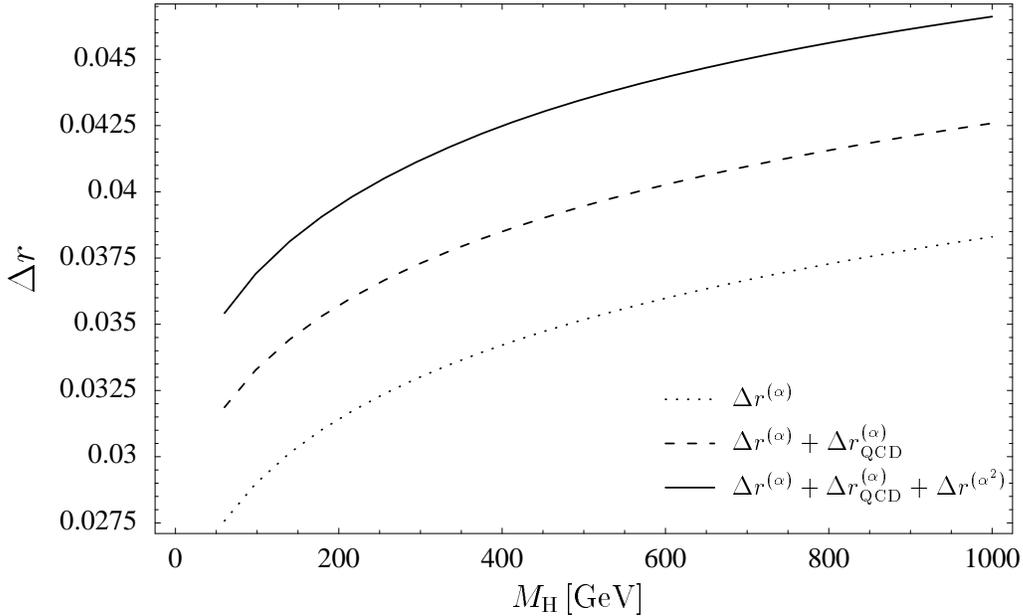,width=14cm}
}
\caption{Contribution of one-loop and higher order corrections to $\Delta r$.}
\label{fig:deltar}
\end{figure}

In the previous sections the characteristics of the calculation of electroweak
two-loop contributions to $\Delta r$ have been pointed out. Combining the
fermionic ${\mathcal{O}}(\alpha^2)$ contributions with the one-loop and the QCD
corrections yields the total result
\begin{eqnarray}
\Delta r = \Delta r^{(\alpha)} + \Delta r^{(\alpha\alpha_s)} +
  \Delta r^{(\alpha\alpha_s^2)} +
  \Delta r^{(N_f \alpha^2)} + \Delta r^{(N_f^2 \alpha^2)}. \label{drtotal}
\end{eqnarray}
Here $N_f, N_f^2$ symbolize one and two fermionic loops respectively. Fig.
\ref{fig:deltar} shows that both the QCD and electroweak two-loop corrections
give sizeable contributions of 10--15\% with respect to the one-loop result.

\begin{figure}[tb]
\centerline{
\psfig{figure=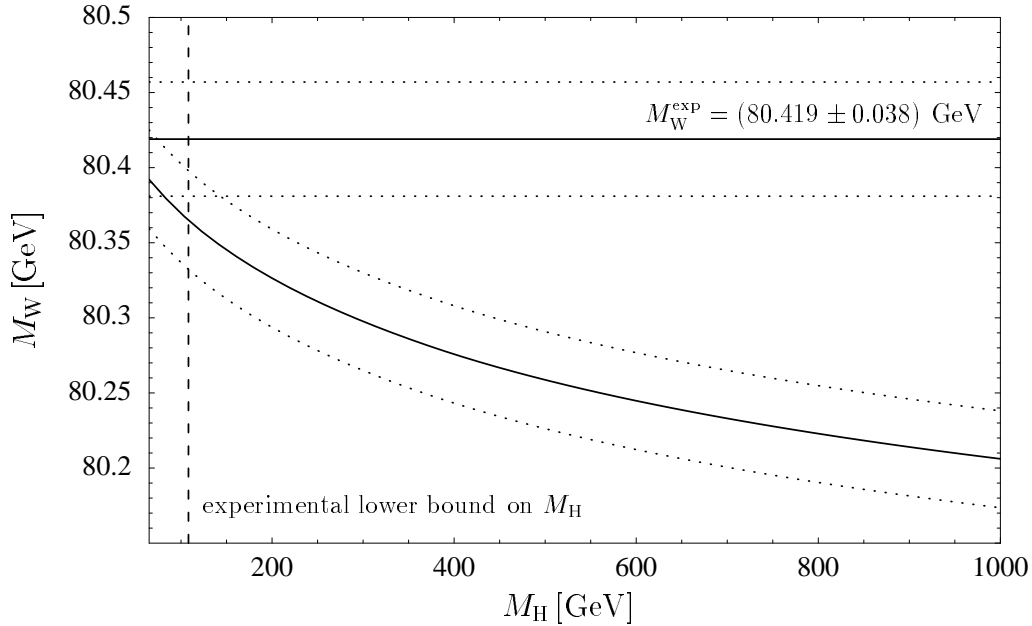,width=14cm}
}
\caption{Prediction for $\MW$ as function of $\MH$ compared with the
experimental W mass.}
\label{fig:MWpred}
\end{figure}

In Fig. \ref{fig:MWpred} the prediction for $\MW$ derived from the result
\eqref{drtotal} and the relation \eqref{MWpred} is compared with the
experimental value for $\MW$. Dotted lines indicate one standard deviation
bounds. The main uncertainties of the prediction originate from the
experimental errors of $\mt = (174.3 \pm 5.1)$ GeV \cite{C:pdg} and
$\Delta\alpha = 0.05954 \pm 0.00065$ \cite{dalp}. It can be noted that light
Higgs masses are favored by this analysis. Further implications of the
precision calculation of $\MW$ are discussed in \cite{G:talk}.

These results can be compared with the results obtained by expansion of the
two-loop contributions up to next-to-leading order in $\mt$ \cite{mtsqr,mtsqr2}. The predicted values
for $\MW$ for several values of $\MH$ are given in Tab. \ref{tab:MWcomp}.
\begin{table}[b]
\caption{Comparison between $\MW$--predictions from a NLO expansion in $\mt$
($\MW^{\stxt{expa}}$) and the full calculation ($\MW^{\stxt{full}}$). $\delta
\MW$ denotes the difference. Experimental
input values are taken from \cite{mtsqr2}.}
\label{tab:MWcomp}
\setlength{\tabcolsep}{1.1pc}
\newlength{\digitwidth} \settowidth{\digitwidth}{\rm 0}
\catcode`?=\active \def?{\kern\digitwidth}
\begin{tabular*}{\columnwidth}{@{}r@{\extracolsep\fill}r@{\extracolsep\fill}
        r@{\extracolsep\fill}r@{\extracolsep\fill}r}
\hline
\rule{0mm}{1em}$\MH$ &
$\MW^{\stxt{expa}}$ &
$\MW^{\stxt{full}}$ &
$\delta \MW$ \\
\rule{0mm}{0mm} [GeV] &
[GeV] &
[GeV] &
[MeV] \\
\hline
$65$ & $80.4039$ & $80.3997$ & $4.2$  \\
$100$ & $80.3805$ & $80.3771$ & $3.4$ \\
$300$ & $80.3061$ & $80.3051$ & $1.0$ \\
$600$ & $80.2521$ & $80.2521$ & $0.0$ \\
$1000$& $80.2129$ & $80.2134$ & $-0.5$ \\
\hline
\end{tabular*}
\end{table}
Agreement is found between the results with maximal 
deviations of about 4 MeV in $\MW$.

The theoretical uncertainty due to missing higher order contributions can be
estimated as follows. The missing ${\mathcal{O}}(\alpha^2)$ purely bosonic
corrections can be judged by means of resummation relations to be very small
($< 1$ MeV effect on $\MW$ for a light Higgs). An estimate of the
${\mathcal{O}}(\alpha^3)$ terms can be obtained from the
renormalization scheme dependence of the two-loop result yielding about 2--3
MeV, and the missing higher order QCD corrections were estimated to be about
4--5 MeV \cite{QCDmiss,DG:error}. Adding this up linearly, one arrives at a
total
uncertainty of about 7 MeV for the $\MW$--prediction at low Higgs masses ($\MH
\lesim 150$ GeV).

\section{Conclusion}

In this talk the realization of an exact two-loop calculation of fermionic
contributions in the full electroweak SM and its application to the
precise computation of $\Delta r$ were presented. Numerical results and an
estimate of the remaining uncertainties were given and might serve as
ingredient for future SM fits.

\vspace{1em}
The speaker would like to thank S. Bauberger and D. St\"ockinger for valuable
discussions.

\end{document}